\documentclass[%
 reprint,
superscriptaddress,
 amsmath,amssymb,
 aps,
prl,
]{revtex4-2}

\usepackage{graphicx}
\usepackage{dcolumn}
\usepackage{bm}

\usepackage{esvect}

\begin{document}


\title{Deeply Sub-Wavelength Localization with Reverberation-Coded-Aperture}

\author{Michael del Hougne}
\affiliation{Julius-Maximilians-Universität Würzburg, D-97074 Würzburg, Germany}%
\author{Sylvain Gigan}%
\affiliation{Laboratoire Kastler Brossel, Université Pierre et Marie Curie,
Ecole Normale Supérieure, CNRS, Collège de France, F-75005 Paris, France}%

\author{Philipp del Hougne}%
 \email{philipp.delhougne@gmail.com}
\affiliation{Univ Rennes, CNRS, IETR - UMR 6164, F-35000, Rennes, France}%

\begin{abstract}

Accessing sub-wavelength information about a scene from the far-field without invasive near-field manipulations is a fundamental challenge in wave engineering. Yet it is well understood that the dwell time of waves in complex media sets the scale for the waves’ sensitivity to perturbations. Modern coded-aperture imagers leverage the degrees of freedom (DoF) offered by complex media as natural multiplexor but do not recognize and reap the fundamental difference between placing the object of interest \textit{outside} or \textit{within} the complex medium. Here, we show that the precision of localizing a sub-wavelength object can be improved by several orders of magnitude simply by enclosing it in its far field with a reverberant chaotic cavity. We identify deep learning as suitable noise-robust tool to extract sub-wavelength information encoded in multiplexed measurements, achieving resolutions well beyond those available in the training data. We demonstrate our finding in the microwave domain: harnessing the configurational DoF of a simple programmable metasurface, we localize a sub-wavelength object inside a chaotic cavity with a resolution of $\lambda/76$ using intensity-only single-frequency single-pixel measurements. Our results may have important applications in photoacoustic imaging as well as human-machine interaction based on reverberating elastic waves, sound or microwaves.

\end{abstract}

\maketitle

Retrieving a representation of an object based on how it scatters waves is a central goal across all areas of wave engineering (light, microwaves, sound, ...) with applications ranging from biomedicine via microelectronics to astrophysics. Since wave energy cannot, in general, be focused beyond the diffraction limit in free space, a widespread misconception is that sub-wavelength information can \textit{only} be accessed via evanescent waves. This argument ignores the crucial roles of \textit{a priori} knowledge and signal-to-noise ratio (SNR); moreover, many imaging schemes do not even rely on focusing. Indeed, given extensive \textit{a priori} knowledge, an imaging task can collapse to a curve fitting exercise without any fundamental bound on the achievable precision (e.g.~deconvolution microscopy~\cite{sibarita2005deconvolution,sarder2006deconvolution}). The advent of deep learning has enabled elaborate demonstrations of such nonlinear function approximations, facilitating deeply sub-wavelength imaging even with a simple plane wave from the far field~\cite{pu2020label}.
Despite the resulting frequent absence of any wavelength-induced fundamental resolution bounds, specific physical mechanisms can be useful to boost the practically achievable resolution. A common example is the above-mentioned access to evanescent waves either via near-field measurements~\cite{betzig1993single,seo2020near} or by coupling them to the far-field with near-field scatterers~\cite{pendry2000negative,fang2005sub,guenneau2007acoustic,sukhovich2009experimental,lemoult2010resonant,van2011scattering,zhu2011holey,park2013subwavelength,park2014full,baha}. Similarly to the use of fluorescent markers~\cite{hell1994breaking}, these approaches are inherently invasive since they rely on manipulations of the object's near-field. A further notable idea relies on tailored coherent far-field illumination to create super-oscillatory hotspots~\cite{rogers2012super,dubois2015time} but suffers from inherently low SNRs.

In the wave chaos community~\cite{stockmann2000quantum}, it is well known that a wave's sensitivity to geometrical perturbations~\cite{gorin2006dynamics,kuhl2016microwave,taddese2010sensing,taddese2013quantifying} is directly related to its dwell time in the interaction domain~\cite{brouwer1997quantum}. 
This effect can be thought of as a generalization of the sub-wavelength interferometric sensitivity in phase microscopy~\cite{park2018quantitative,juffmann2020local,bouchet2020fundamental} or high-finesse Fabry-Perot cavities~\cite{arcizet2006high}. 
If the scene to be imaged is enclosed in its far field by a reverberant chaotic cavity, the dwell time is drastically enhanced. Different scenes can then be interpreted as different perturbations of an otherwise static complex scattering geometry. This clearly hints at the potential of chaotic reverberation to significantly lower the resolution limit without any near-field manipulation -- provided that the complete scrambling of waves (and the information that they carry) inside the complex medium can be untangled in post-processing. 

While linear chaotic reverberation as simple route to deeply sub-wavelength resolution has to date remained unexplored, possibly with the exception of diffusing wave spectroscopy capable of extracting \textit{global} features (e.g., scattering cross-section) of \textit{moving} scatterers in complex media~\cite{pine1988diffusing,maret1997diffusing,de2003field,taddese2013quantifying}, a rich literature actually exists on imaging and sensing with a complex medium as coded aperture (CA). This research track is driven by the desire to achieve imaging with as few detectors and measurements as possible. Rather than directly mapping the object to its image, the spatial object information can be multiplexed across random configurations of a CA onto a single detector~\cite{chan2008single} -- see Fig.~1(a). Practical implementations of CAs often leverage the fact that wave transmission \textit{through} a complex medium (multiply scattering medium, chaotic cavity, disordered metamaterial) constitutes random multiplexing thanks to the medium’s spectral, spatial or configurational degrees of freedom (DoF)~\cite{liutkus2014imaging,hunt2013metamaterial,fromenteze2015computational,xie2015single,sleasman2015dynamic} -- Fig.~1(b) illustrates the former. In other words, the transmission matrix of a complex medium naturally offers the desired properties of a random multiplexing matrix~\cite{liutkus2014imaging,fromenteze2015computational}.  

Within this realm, scenarios in which object and wave source are embedded \textit{within} the complex medium, as in Fig.~1(c), have been treated as a simple alternative way of natural random multiplexing. In photoacoustics, it was recently suggested to enclose the imaging target in an acoustically reverberant cavity~\cite{brown2019reverberant}. Moreover, several schemes for human-machine interaction are inevitably confronted with waves reverberating around an object, for instance, object localization with microwaves in indoor environments or with elastic waves in solid plates~\cite{6G_WhitePaper,del2018precise,ing2005solid,liu2009tactile}. However, the benefits of such ``reverberation-coded-apertures'' (RCAs) go far beyond randomized multiplexing. If the object is inside (rather than outside) the complex medium, the wave interacts with the object not once but countless times, thereby developing a much higher sensitivity to sub-wavelength object details. 
Recent efforts to construct optimal coherent states for sensing in complex media~\cite{bouchet2020influence,bouchet2021maximum} differ from our problem, besides their requirement for multi-channel excitation, in that they rely on (and are limited to) small perturbations of the sought-after variable.

In this Letter, we introduce RCAs, combined with deep learning to extract encoded sub-wavelength information, as truly non-invasive route to deeply sub-wavelength resolution. Previous works on setups that can be considered as RCAs assumed that resolution was inherently diffraction-limited~\cite{del2018precise,ing2005solid}, or that a non-linear process resulting in self-oscillations was required for sub-wavelength resolution~\cite{cohen2011subwavelength}. Here, we first establish the fundamental link between dwell time, wave sensitivity and resolution in a semi-analytical study of the prototypical example of object localization inside a linear chaotic cavity using spectral DoF. Then, we demonstrate our finding experimentally in the microwave domain, using configurational DoF provided by a simple programmable metasurface.

\begin{figure}
\centering
\includegraphics [width =  \columnwidth]{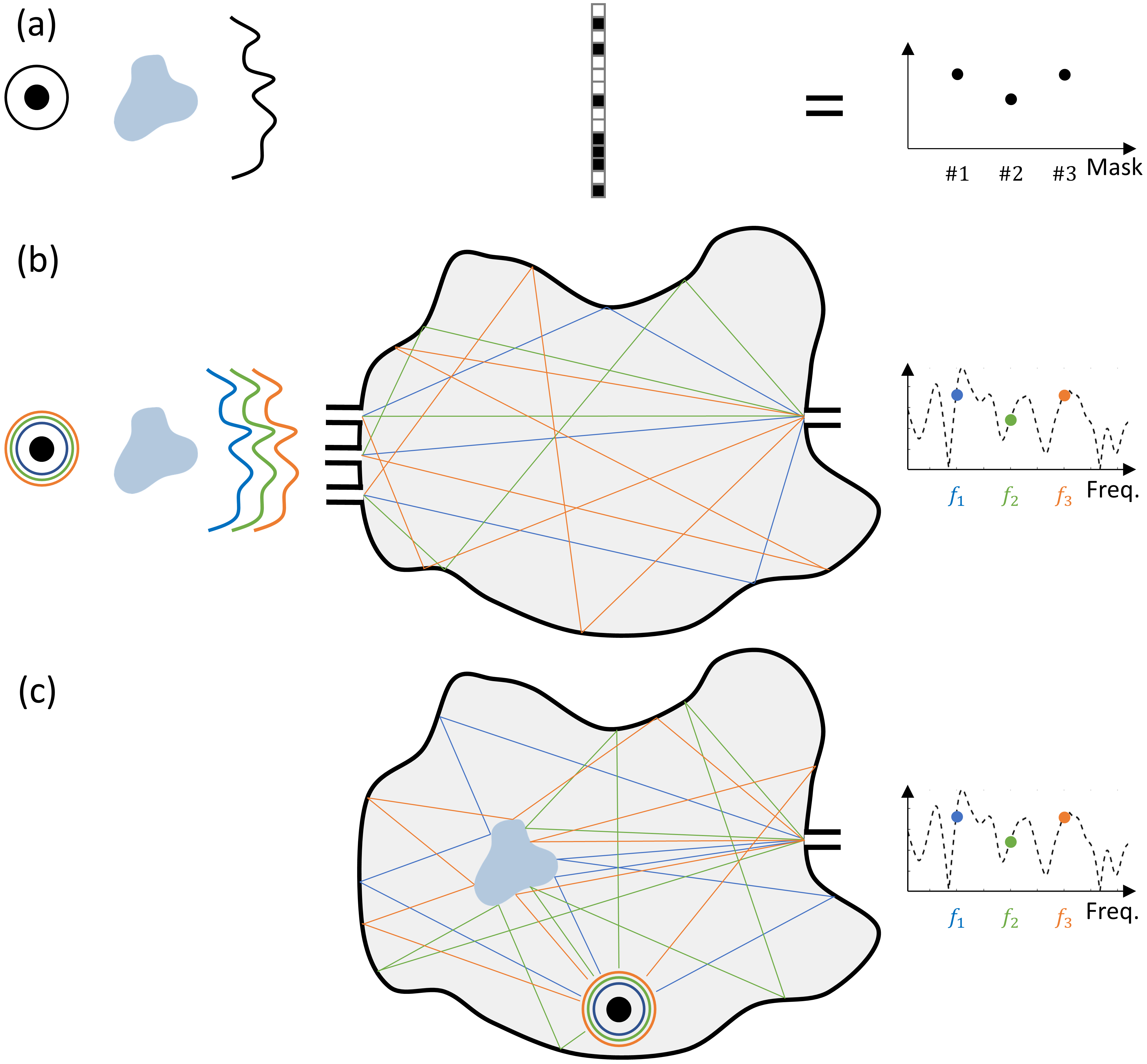}
\caption{(a) Conventional CA. A wavefront is scattered by an object and then multiplexed across different masks onto a single-pixel detector. (b) Use of a complex medium's spectral DoF (color-coded) as CA. Wavefronts are scattered by an object and then propagate \textit{through} a chaotic cavity such that spatial information is multiplexed across different frequencies captured by a single-pixel detector. (c) Reverberation-coded-aperture: same as (b) except that object and wave source are \textit{inside} the chaotic cavity. 
}
\label{fig1}
\end{figure}

To start, we consider for concreteness a 2D model problem in semi-analytical simulations based on a coupled-dipole formalism (see Refs.~\cite{baha,localiz_dynamic_environm} and SM). As depicted in Fig.~2(a-c), we consider three scenarios expected to correspond to different durations of the scattering process: free space, a cavity with quality factor $Q=263$ and a cavity with $Q=556$. In each case, the transmission $S_{12}$ between a transmit and a receive port is measured for various frequencies in order to localize (using ANN-based data analysis) a non-resonant dipole that could be located anywhere along a circular perimeter. Given the single-channel nature of our single-detector scheme, we estimate the duration of the scattering process via the ``phase delay time'' $\tau = \partial \text{arg}(S_{12})/\partial \omega$~\cite{RevModPhys.89.015005}. 
The irregularly shaped cavity constitutes a complex medium for wave propagation in which $\tau$ is hence a statistically distributed quantity; in Fig.~2(d), we plot the cumulative distribution function (CDF) of its magnitude for the three considered cases, confirming that they correspond to increasing dwell times. Compared to a regular cavity, such a chaotic cavity has not only the practical advantage of being easy to implement but also that ergodicity ensures statistically similar properties~\cite{hemmady2005universalPRL,hemmady2005universal} irrespective of the object position.

\begin{figure}
\centering
\includegraphics [width =  \columnwidth]{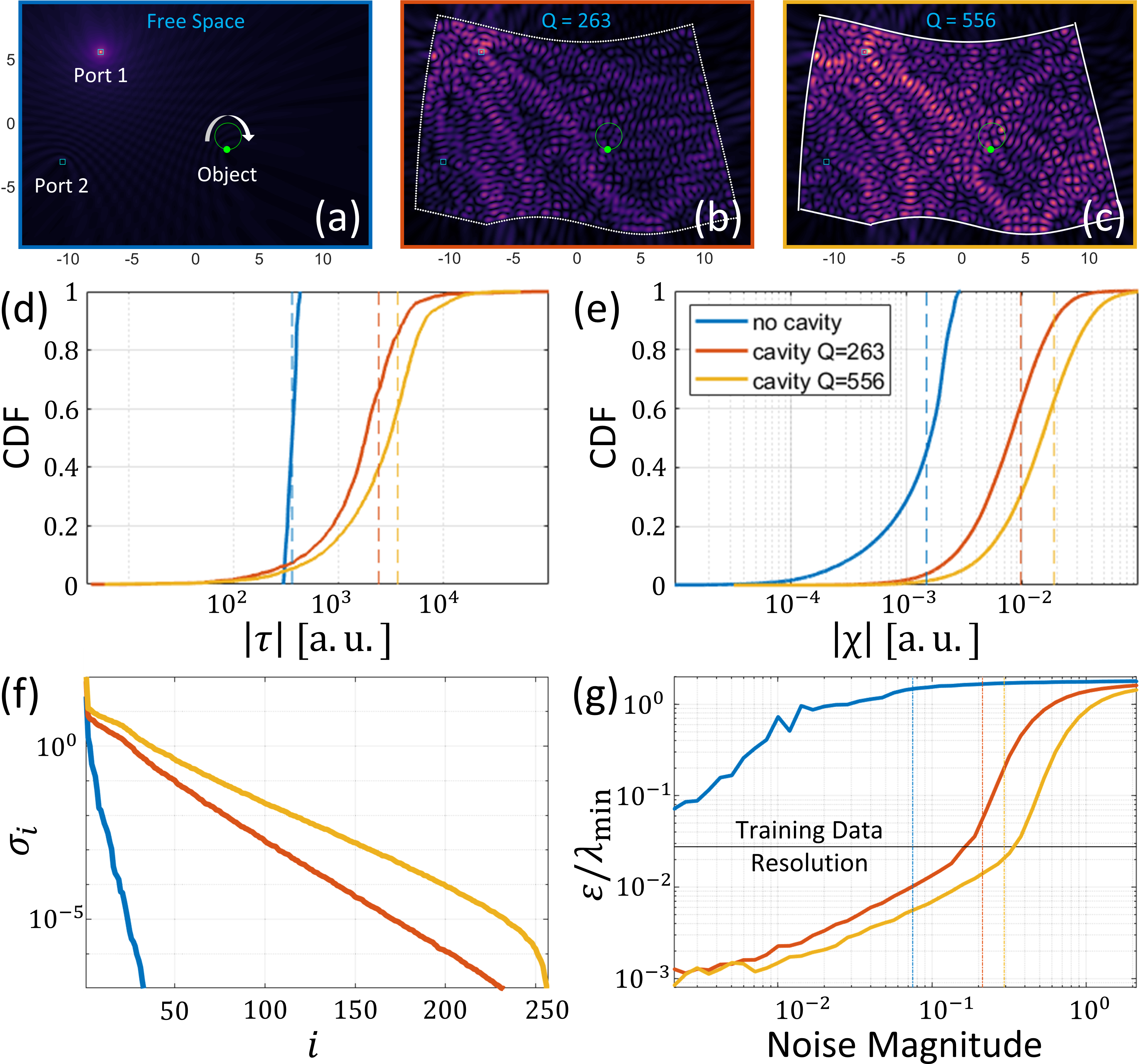}
\caption{(a-c) Electric field magnitude in 2D semi-analytical simulations for (a) free space, a chaotic cavity with (b) $Q=263$ and (c) $Q=556$. All maps use the same color scale. 
(d) CDF of the dwell time magnitude $\left|\tau\right|$ distribution in the three cases. Vertical dashed lines indicate the corresponding mean values. 
(e) CDF of the parametric derivative magnitude $\left|\chi\right|$ (with respect to the object position) in the three cases. Vertical dashed lines indicate the corresponding mean values. 
(f) SV spectra of $\textbf{T}(X,f)$ (without any noise) for the three considered cases. 
(g) Average localization error $\epsilon$ in terms of the smallest utilized wavelength $\lambda_{\text{min}}$ as a function of the \textit{absolute} magnitude of the measurement noise for the three cases. Vertical dashed lines indicate the corresponding signal magnitudes. The horizontal black line indicates the training data resolution.}
\label{fig2}
\end{figure}

The dwell time plays a crucial role in mesoscopic physics because it is related to several other relevant quantities~\cite{genack1999statistics,davy2015transmission,savo2017observation,RevModPhys.89.015005}, some of which happen to also be critical metrics for RCA-based imaging and sensing. The most obvious quantity is the energy stored in the complex medium~\cite{RevModPhys.89.015005,durand2019optimizing,del2020experimental} which is directly related to the received signal strength and thereby the measurement's SNR (assuming detector-induced noise). 
Of course, measurements with higher SNR contain more information.
The direct link between $|\tau|$ and the stored energy is apparent upon visual inspection of Fig.~2(a-c), and indeed the received signal strength is on average 2.9 [3.9] times higher in (b) [(c)] than in (a). 

The sensitivity to parametric perturbations of the scattering system is another important quantity that can be related to the wave's dwell time in the interaction domain~\cite{brouwer1997quantum,del2020demand}. Intuitively, this can be understood as follows: the probability that a tiny perturbation impacts the evolution of a wave increases with the wave's lifetime in the interaction domain. Analogous to $\tau$, we define $\chi = \partial S_{12}/\partial X$, where $X$ denotes the considered parameter (the object position along the allowed trajectory in our case). Indeed, in Fig.~2(d,e) we observe a clear correspondence between the distributions of $|\tau|$ and $|\chi|$, for instance in terms of the mean value (dashed vertical lines). The difference between two nearby object positions in terms of the corresponding scattering matrices (specifically, $S_{12}$) is thus on average larger if the dwell time in the interaction domain is larger. This effect is induced by using a RCA instead of a conventional CA for which the object scatters the wave before the wave is multiplexed across the CA.

In the case of a CA (RCA or conventional) leveraging spectral DoF, the amount of information that can be extracted from measurements within a given bandwidth is also tightly linked to the dwell time. Indeed, the spectral decorrelation is related to the rate at which $S_{12}$ fluctuates with respect to the frequency. The lower the correlation between different measurement modes is, the less redundant information is acquired. To illustrate this effect, we consider the singular value (SV) decomposition of a 2D matrix $\textbf{T}(X,f)$ containing the measured transmission for different frequencies and positions along the allowed trajectory; $\textbf{T}(X,f)$ later serves as training data for the ANN. In Fig.~2(f), we plot the SV spectra of this matrix 
for all three considered cases. Indeed, higher dwell times correlate with a flatter SV spectrum, implying that different frequencies are less correlated.

Having established three distinct RCA mechanisms expected to enhance the possibility of \textit{physically encoding} deeply sub-wavelength information about the object via wave propagation inside the RCA in multiplexed measurements, we now tackle the problem of \textit{digitally decoding} this information. In order to approximate an inverse function of the physical wave scattering process, mapping the measured data to the object position, we use deep learning. 
We deliberately use an artificial neural network (ANN) consisting of several fully connected layers (see SM
) as opposed to more popular convolutional 
architectures because the latter excel at identifying relevant local correlations in the data whereas we hypothesize that the complete scrambling caused by wave scattering encodes the relevant features in long-range correlations within the data~\cite{zheng2018processing,localiz_dynamic_environm,zhu2020image}. Moreover, our ANN does not solve a classification problem but predicts a continuous variable: the object's position. 

We report in Fig.~2(g) the average localization error in terms of the smallest utilized wavelength, $\epsilon/\lambda_{\text{min}}$, as a function of the measurement noise magnitude. 
First, we observe that even the free space scenario can achieve deeply sub-wavelength resolution beyond $\lambda_{\text{min}}/10$ at low noise levels, stressing the absence of any fundamental wavelength-induced resolution bounds, similar to Refs.~\cite{sibarita2005deconvolution,sarder2006deconvolution,pu2020label}. 
Second, as hypothesized, the longer the dwell time in the RCA, the higher the achievable resolution at a given noise level. In our case, we observe resolutions beyond $\lambda_{\text{min}}/10^3$ but, as justified above, we refrain from comparing these absolute resolution values to other works with different \textit{a priori} knowledge and SNR.
Third, remarkably, the achievable resolution can be more than an order of magnitude better than the resolution of the training data, suggesting that beyond being an efficient approximator to ``seen'' data, our ANN also very faithfully interpolates between ``seen'' data points. 
Deep learning also offers a remarkable noise-robustness which significantly outperforms a simple multivariate linear regression (see SM section D and Fig. S3).

One question naturally arises: can we isolate the contribution of the three identified RCA mechanisms? Specifically, we now evidence the major role of the dwell-time-enhanced sensitivity to tiny perturbations. To that end, we consider a setting in which the other two factors do not impact the localization accuracy: we compare the three considered scenarios in terms of their SNR (removing benefits due to enhanced signal strength), and operate with a single DoF (removing benefits due to faster spectral decorrelation). 
Table~1 summarizes the achievable localization accuracies for two rather high values of SNR; the focus here is not on the absolute localization precision but on how it compares between the three considered scenarios. 
The dependence of the achieved localization accuracy on the dwell time emerges very clearly, confirming our argument that placing the scene inside a RCA rather than outside a CA boosts the sensitivity to sub-wavelength scene details. Incidentally, even in Table~1 we observe a resolution of $\sim \lambda/3.8$ for the RCA with $Q=556$ at an SNR of 30~dB.

\begin{table}[t]
\caption{\label{table_singleDoF} Average localization error $\epsilon / \lambda_{\text{min}}$ using a single DoF for two magnitudes of the measurement noise \textit{relative} to the measured signal strength.}
\begin{ruledtabular}
\begin{tabular}{cccccc}
SNR [dB] &  no cavity & cavity $Q=263$ & cavity $Q=556$\\
\hline
30 & 1.12 & 0.53  & 0.26 \\
60 & 0.59 & 0.31 & 0.24 \\

\end{tabular}
\end{ruledtabular}
\end{table}

Having investigated fundamental RCA mechanisms, we now report an experimental demonstration in the microwave domain inside a 3D irregularly shaped metallic enclosure -- see Fig.~3(a). We use configurational instead of spectral DoF: we measure the transmission between two antennas (``single-pixel detector'') at a single frequency ($f_0 = 2.463\ \text{GHz}$) but for 
a fixed series of random 
configurations (parameter $c$) of the cavity's scattering properties. The latter is conveniently implemented with a simple programmable metasurface~\cite{cui2014coding,kaina2014hybridized} consisting of an array of individually tunable meta-atoms with two digitalized states mimicking Dirichlet or Neumann boundary conditions~\cite{dupre2015wave}. Moreover, we now only use the intensity information of the measurements, to illustrate that RCAs enable deeply sub-wavelength resolution even without access to phase information, which relaxes hardware requirements considerably. We access different dwell time regimes by tuning the opening of the cavity's ceiling.

\begin{figure}
\centering
\includegraphics [width =  \columnwidth]{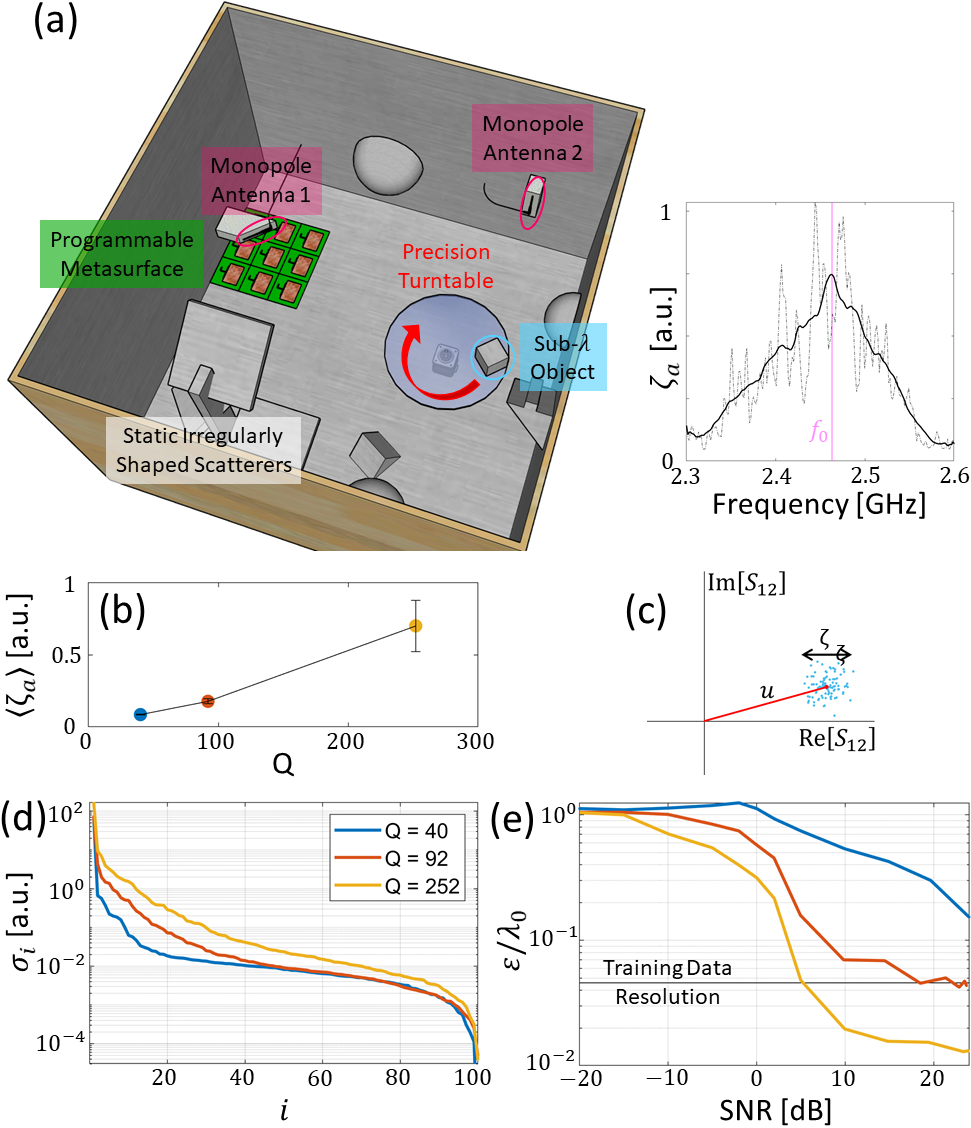}
\caption{(a) Experimental setup: a 3D complex scattering enclosure contains a sub-wavelength metallic object on a precision turntable, two monopole antennas, and a programmable metasurface in the vicinity of one antenna. The enclosure's ceiling can be open ($Q=40$), partially covered ($Q=92$) or fully covered ($Q=252$) with metal. See SM for technical details. The inset shows the standard deviation $\zeta_a$ of $|S_{12}|$ over 100 random metasurface configurations and identifies the chosen operating frequency $f_0$. 
(b) Dependence of mean of $\zeta_a$ over object positions on $Q$. 
(c) Sketch of $S_{12}(f_0)$ distribution for 100 random metasurface configurations. 
(d) SV spectra of $\textbf{T}(X,c)$ for the three considered cases. 
(e) Average localization error $\epsilon$ in terms of the utilized wavelength $\lambda_0$ as a function of SNR.  The horizontal black line indicates the training data resolution.}
\label{fig3}
\end{figure}

Of the above identified three distinct RCA mechanisms, two (signal strength and sensitivity) are independent of the utilized type of DoF; the third (measurement diversity) turns out to be favorably linked to enhanced dwell times if configurational DoF are used, too, albeit for a different reason. The amount of information that can be extracted from a series of measurements at $f_0$ with random metasurface configurations is larger if the latter induce stronger fluctuations of $S_{12}(f_0)$. A longer dwell time correlates with a larger standard deviation of $|S_{12}(f_0)|$, as evidenced in Fig.~3(b). 
This can be understood by decomposing the transmission between the two ports into all contributing ray paths. If the metasurface is small compared to the cavity surface and the dwell time is relatively low, only a few rays are affected by the metasurface configuration. We sketch for such a scenario the cloud of accessible $S_{12}(f_0)$ values in the Argand diagram in Fig.~3(c); the cloud is not centered on the origin because many rays are not controlled by the metasurface. The longer the dwell time, the more rays will encounter some of the metasurface elements such that the radius of the cloud increases, as witnessed in Fig.~3(b). To illustrate that $\textbf{T}(X,c)$ contains more information if the dwell time is longer, we plot the corresponding SV spectra in Fig.~3(d). A qualitatively similar trend as in Fig.~2(f) is seen, despite the use of a different type of DoF.

The average experimental localization error plotted in Fig.~3(e) is consequently significantly lower if the dwell time is longer. The dependence on the SNR is evaluated by adding white noise to the experimentally measured values. Unlike in Fig.~2(g), we plot $\epsilon/\lambda_0$ as a function of the relative rather than absolute noise magnitude due to experimental constraints, such that the curve does not reflect the first RCA mechanism's benefits (signal strength). Despite the use of low-cost measurement equipment (see SM) and intensity-only data, we achieve resolutions up to $\lambda/76$ in our experiment. For $Q=252$, we observe once again that our ANN decoder achieves a resolution clearly exceeding that of the training data.

To conclude, in this Letter, we proved that reverberation in a complex medium efficiently encodes deeply sub-wavelength details in multiplexed measurements without any manipulation of the object's near field. We evidenced that the wave's dwell time is directly linked to the achievable resolution via three mechanisms, irrespective of the utilized type of DoF: (i) enhanced signal strength, (ii) enhanced sensitivity, and (iii) enhanced measurement diversity. We further showed that ANNs are capable of decoding such measurements with unexpectedly high fidelity. In microwave experiments in a chaotic cavity leveraging the configurational DoF offered by a programmable metasurface, we successfully localized sub-wavelength objects with a resolution of $\lambda/76$.

Looking forward, evaluating potential benefits of using a self-oscillating source~\cite{lobkis2009larsen,cohen2011subwavelength} that operates at a real-valued pole associated with a diverging delay time is a first avenue for future exploration. We also envision a ``learned RCA'' that jointly optimizes physical encoding and digital decoding in an end-to-end fashion~\cite{del2020learned}. Moreover, our method can be extended conceptually from object localization to more complex imaging tasks, e.g. to recognize sub-wavelength object shapes inside a RCA without any near-field manipulation. In a more applied context, various single-element detection methods, including RCAs, are currently being explored in photoacoustic imaging~\cite{brown2019reverberant,li2020snapshot,li2020multifocal}; our work encourages investigating whether photoacoustic imaging can capitalize on potential resolution enhancements offered by the RCA but not by other ``ergodic relays''. Further practical applications lie in emerging techniques for human-machine interactions which naturally deal with reverberating waves of different types~\cite{6G_WhitePaper,liu2009tactile}. In the optical domain, our scheme may be applied to the deeply sub-wavelength localization of microscopic particles, by sandwiching them between two multiply-scattering slabs such as white paint or biological tissue~\cite{bertolotti2012non}.

\providecommand{\noopsort}[1]{}\providecommand{\singleletter}[1]{#1}%

\clearpage

\begin{center}
\textbf{SUPPLEMENTAL MATERIAL}
\end{center}

\bigskip

\renewcommand{\thefigure}{S\arabic{figure}}
\renewcommand{\theequation}{S\arabic{equation}}
\setcounter{equation}{0}
\setcounter{figure}{0}

For the interested reader, here we provide numerous additional details that complement the
manuscript and may support any efforts to reproduce our work. This document is organized as
follows:

A. Semi-Analytical Simulations.

B. Experimental Setup.

C. Artificial Neural Network.

D. Comparison of Linear Multi-Variable Regression and Deep Learning.

E. Further Parameters that Impact the Localization Precision.

\maketitle

\subsection{Semi-Analytical Simulations}

We consider a 2D model of $N$ dipoles in the $x-y$ plane whose dipole moments are oriented along the vertical $z$ axis~\cite{baha_B,localiz_dynamic_environm_B}. The dipole moment $p_i$ of the $i$th dipole is related to the local electric field at the dipole’s position $\vv{r_i}$ via the dipole’s polarizability $\alpha_i$:

\begin{equation}
\vv{p_i}(f) = \alpha_i(f) \vv{E_{\text{loc}}}(\vv{r_i},f).
\label{eq:p}
\end{equation}

\noindent A Lorentzian model is used for the inverse polarizability, 

\begin{equation}
\alpha_i^{-1}(f) = 4\pi^2\gamma \left( f_{\text{res}}^2 - f^2 \right) + j \frac{k^2}{4\epsilon},
\label{eq:alpha}
\end{equation}

\noindent where the imaginary part corresponds to radiation damping in accordance with the optical theorem. $k$ denotes the wave vector and $\epsilon$ is the medium's relative permittivity. The local field at the $i$th dipole is the superposition of the external field exciting the system and the fields radiated by the other dipoles: 

\begin{equation}
\vv{E_{\text{loc}}}(\vv{r_i},f) = \vv{E_{\text{ext}}}(\vv{r_i},f) + \sum_{j \neq i}G_{ij}\left( \vv{r_i},\vv{r_j},f\right)\vv{p_j}(f).
\label{eq:Eloc}
\end{equation}

\noindent Here, 

\begin{equation}
G_{ij}\left( \vv{r_i},\vv{r_j},f\right) = -j \frac{k^2}{4\epsilon} \text{H}_0^{(2)}\left( \frac{2 \pi f}{c} \left| \vv{r_i} - \vv{r_j} \right|\right)
\label{eq:G}
\end{equation}

\noindent is the Green’s function between the positions $\vv{r_i}$ and $\vv{r_j}$ with $\text{H}_0^{(2)}(\dots)$ denoting a Hankel function of the second kind. We hence arrive at the relation 

\begin{equation}
\alpha_i^{-1}(f)\vv{p_i}(f) - \sum_{j \neq i}G_{ij}\left( \vv{r_i},\vv{r_j},f\right)\vv{p_j}(f) = \vv{E_{\text{ext}}}(\vv{r_i},f)
\label{eq:equation}
\end{equation}

\noindent which can be solved via matrix inversion at each considered frequency. For our problem, we are interested in measuring the transmission coefficient $S_{21}(f)$ between dipoles 1 (excitation) and 2 (detection). This quantity corresponds to the field measured at port
2 if the excitation field is unity at port 1 and zero elsewhere. The resonance frequency of the dipoles constituting cavity fence and object are chosen well above the highest considered frequency to ensure that the cavity and object properties are not heavily frequency dependent. 

The circle on which the object is allowed to be located (see Fig.~2(a-c) of the main text) has a radius of $\lambda_0$ which is the wavelength corresponding to the central frequency of the considered frequency interval. The width of the considered frequency interval is $\Delta f / f_0 = 0.22$. The cavities seen in Fig.~2(b,c) in the main text are created with a dipole ``fence''~\cite{localiz_dynamic_environm_B} whose density is varied to tune the cavity's quality factors. 

\subsection{Experimental Setup}

Our complex scattering enclosure is a $0.8 \times 0.8 \times 0.5\ \text{m}^3$ metallic enclosure with scattering irregularities inside, as seen in Fig.~3(a) of the main text. 
The sub-wavelength object to be localized, a metallic cube of side length $4.5\ \text{cm}$, is mounted on a turntable whose angular rotation is controlled by a stepper motor (SM-42BYG011-25) in 1600 steps, at a distance of $8.9 \ \text{cm}$ from the turntable's center. The stepper motor is controlled via an EasyDriver~V4.5 stepper motor driver and an Arduino microcontroller. 
We measure the field transmitted between two standard WiFi antennas (ANT-24G-HL90-SMA) using a full-duplex two-port software-defined radio (LimeSDR Mini) at 2.463~GHz. The SDR is cooled with a standard CPU fan to keep its temperature constant throughout the experiment. The data acquisition software is based on Ref.~\cite{Lime_B}; as mentioned in the main text, only amplitude information is used to analyze the experimental data.

The programmable metasurface consists of a $3 \times 3$ array of 1-bit reflection-programmable meta-atoms following the design presented in Ref.~\cite{kaina2014hybridized_B}. In short, each meta-atom contains two hybridizing resonators, and the resonance of one of those resonators can be tuned by changing its electrical length via the DC bias voltage of an embedded p-i-n diode. Thereby, the overall meta-atom response at the working frequency is on or off resonance, such that waves with one specific polarization under normal incidence experience a phase-shift of roughly 0 or $\pi$ upon reflection, depending on how the meta-atom is programmed. Inside the complex scattering enclosure, waves are incident from random angles and with random polarizations. Therefore, the specific electromagnetic response of the meta-atom is not important here and our technique leveraging configurational diversity can be implemented with any meta-atom design as long as the meta-atom has at least two digitalized states with distinct electromagnetic responses.
The metasurface is placed not too far from one of the monopole antennas to ensure a strong impact on the transmission despite using only 9 programmable meta-atoms. An Arduino microcontroller imposes the desired bias voltages on each meta-atom's p-i-n diode.

\subsection{Artificial Neural Network}

\subsubsection{Overview of Information Flow}

\begin{figure} [b]
\centering
\includegraphics [width =  \columnwidth]{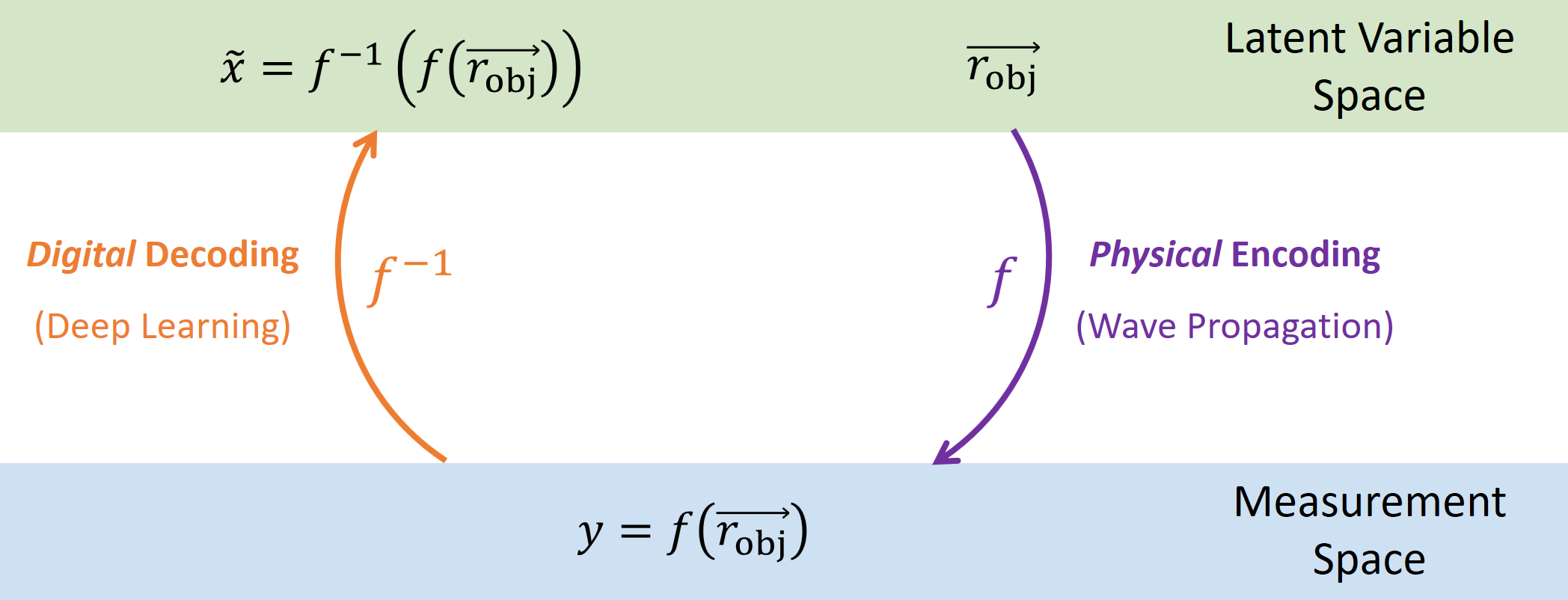}
\caption{Schematic of the flow of information through the physical and digital layers. The latent variable of interest, namely the object position, is first encoded through wave propagation on the physical layer in the measurement space. Then, an ANN is used to approximate an inverse function to return to the latent variable space in order to estimate the object position from the measurements.}
\label{figSA}
\end{figure}

The flow of information in our RCA scheme is summarized in Fig.~\ref{figSA}~\cite{localiz_dynamic_environm_B}. As in any coded-aperture imager, the desired information (here the object position) is not directly measured. Instead, wave propagation acts as a physical encoder and we measure a variable $y$ that is related to the latent variable of interest via a function $f$ that describes the wave propagation. In order to extract the desired information from the measurement, a post-processing step is necessary in which the measurement is decoded digitally, in order to return from the measurement space to the latent variable space. The digital decoder attempts to approximate the inverse function of $f$. Countless approaches to identifying a suitable decoder $f^{-1}$ are in principle conceivable (see also Section D). In the present work, we train a simple fully-connected ANN to approximate $f^{-1}$. 

\subsubsection{ANN Architecture}

\begin{figure}
\centering
\includegraphics [width =  \columnwidth]{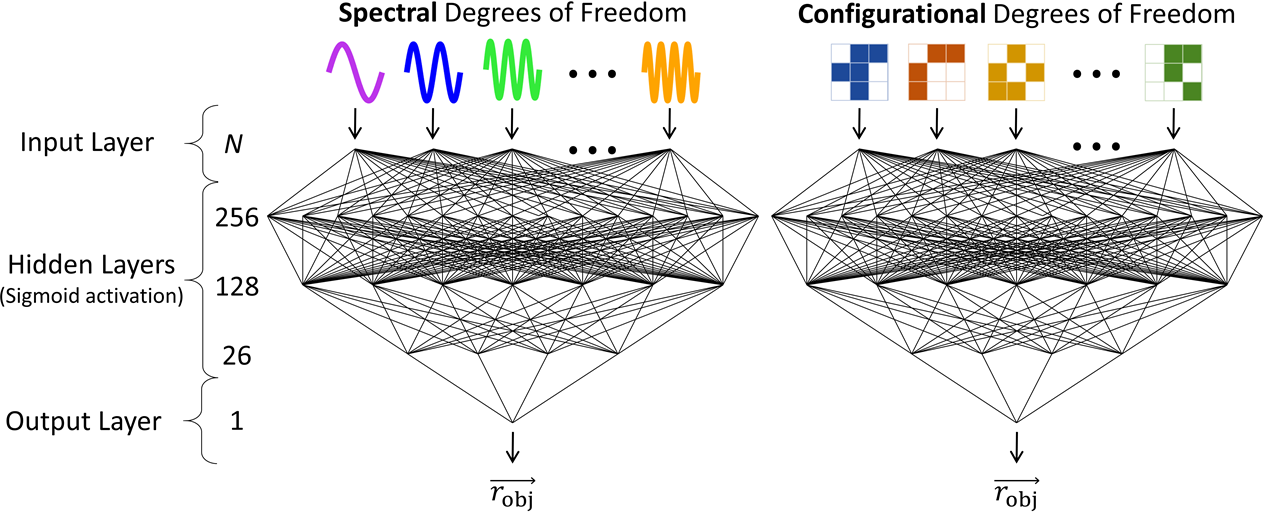}
\caption{Illustration of utilized ANN architecture. $N$ measurements corresponding to the use of $N$ DoF are made: if spectral DoF are used, each measurement is taken at a different frequency (left); if configurational DoF are used, each measurement is taken at the same frequency but for a different metasurface configuration (right). These measurements are then injected into the ANN's input layer (stacking real and imaginary components of the measurements in case magnitude and phase information is available). Multiple hidden layers with sigmoid activation process the information; a single output neuron without activation predicts the object position. Note that the output variable is continuous (this is not a classification ANN).}
\label{figS1}
\end{figure}

Our ANN consists of four fully connected layers with 256, 128, 26 and 2 neurons, respectively, as shown in Fig.~\ref{figS1}. We use the sigmoid function as activation between different layers. No activation function is used after the last layer. In the case of complex valued inputs, we stack real and imaginary components of our data. We normalize the input such that it has zero mean and unity variance (normalization parameters are based on the training data). We train the ANN’s weights via error backpropagation using the Adam optimizer~\cite{adam_B} with a step size of $10^{-3}$. The loss function is defined as the average localization error. We observed that the results did not significantly depend on the exact choice of hyper-parameters (number of layers and neurons); using the root-mean-square localization error instead of the mean error also did not result in significant differences. We underline that this ANN architecture differs in important ways (activation function, loss function) from ANNs used for classification (e.g. in Ref.~\cite{localiz_dynamic_environm_B}) since we intend to predict a continuous variable here. 

Note that our deep learning strategy is supervised, it relies hence on ``labelled'' training data (a series of transmission measurements for which the corresponding object position is known). This terminology potentially causes an unfortunate confusion with the use of ``unlabeled'' in the imaging literature where it refers to the absence of near-field manipulations seeking to label the object of interest with a marker (e.g. a fluorophore).

\subsubsection{Training and Test Data}

For the localization results presented in Fig.~2(g) of the main text based on the semi-analytical simulations, the training data is a $257\times257$ matrix of complex-valued transmission measurements between the two ports. The first dimension corresponds to 257 equally spaced frequencies within a fixed frequency interval, the second dimension corresponds to 257 equally spaced object locations covering the entire allowed circular trajectory. The raw simulation results can be considered noiseless (negligible numerical errors) but zero-mean white Gaussian noise of appropriate standard deviation is added in TensorFlow before the data enters the ANN. Specifically, this means that the noise realization is different in every iteration of training the ANN. The test data is a $257 \times 1000$ matrix corresponding to the same 257 equally spaced frequency points and 1000 completely random positions along the allowed trajectory. For the localization results presented in Fig.~3(e) of the main text based on the microwave-domain experiments, the training data is a $100\times100$ matrix of measured transmission magnitudes between the two ports at the working frequency of 2.463~GHz. The first dimension corresponds to 100 predefined random configurations of the programmable metasurface, the second dimension corresponds to 100 equally spaced object locations covering the entire allowed circular trajectory. 
One fifth of the test data is used as validation data to determine at what epoch the ANN training is stopped (to avoid overfitting), the remaining four fifth are used to compute the reported accuracies.

\subsection{Comparison of Linear Multi-Variable Regression and Deep Learning}

Given the flow of information as illustrated in Fig.~\ref{figSA}, a natural question arises: which digital decoding method should be used? The goal of our present work is not to identify the \textit{best} decoder; instead we claim that our fully-connected ANN from Fig.~\ref{figS1} is a \textit{good} choice of decoder that allows us to prove that useful deeply sub-wavelength information can be extracted from RCA-multiplexed measurements. A thorough discussion of different decoding methods is a signal-processing topic beyond the scope of our present paper in which we focus on the physics behind the links between dwell time, sensitivity and localization precision. The interested reader may refer to Ref.~\cite{localiz_dynamic_environm_B} for a comparison of different decoding methods in a related context (however, Ref.~\cite{localiz_dynamic_environm_B} does not consider the extraction of sub-wavelength information).

Nonetheless, given the apparent simplicity of the sensing task we consider (localizing a single scatterer as opposed to recognizing more complex scattering structures), it may be tempting to assume that the use of deep learning is not really necessary here. While we do not claim that elaborate signal processing methods other than deep learning cannot excel at our problem, we illustrate in this section that a simple linear multi-variable regression does not. 

\begin{figure}
\centering
\includegraphics [width =  \columnwidth]{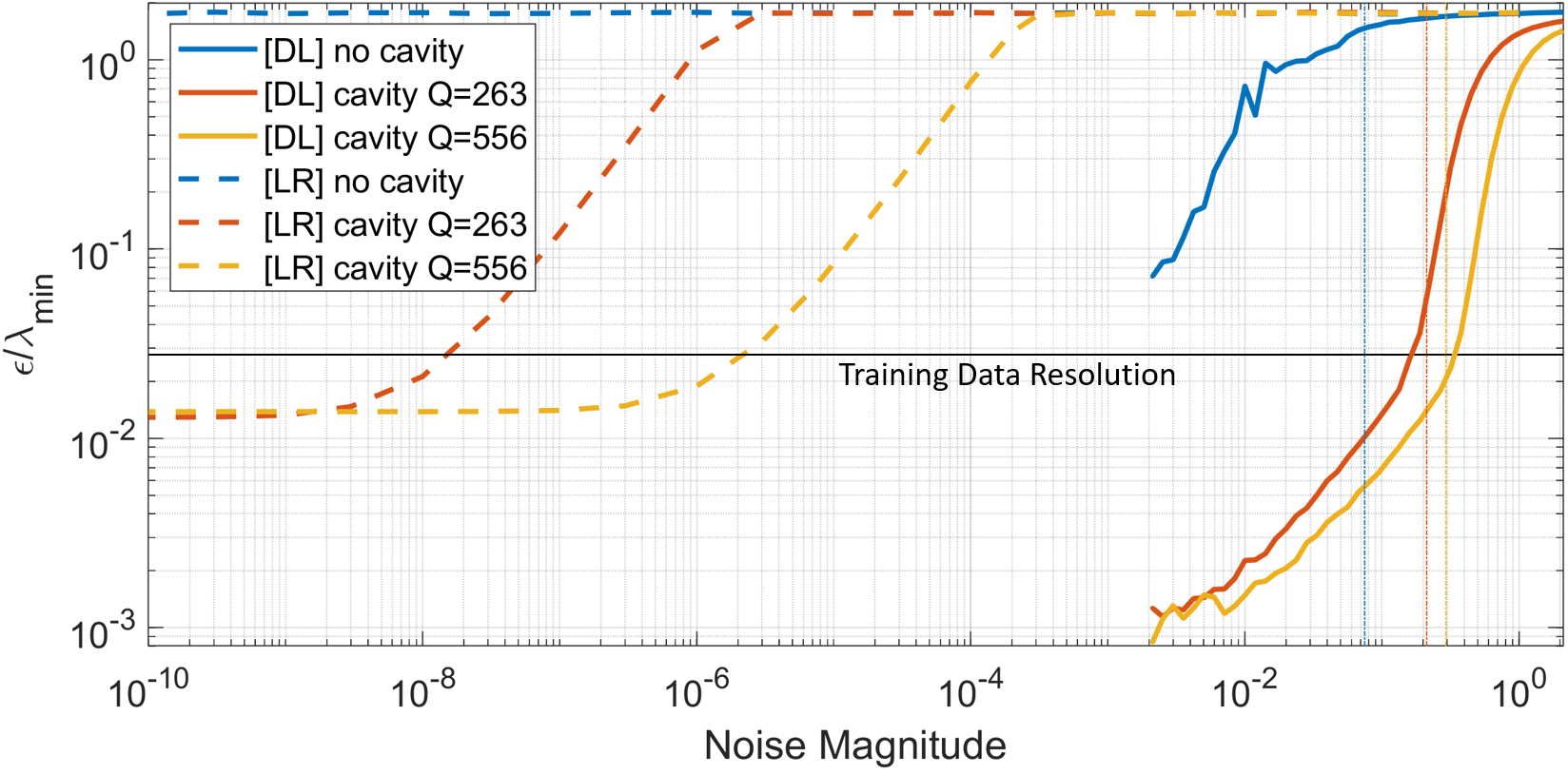}
\caption{Comparison of deep learning (DL, continuous lines) with linear multi-variable regression (LR, dashed lines) as digital decoder of the multiplexed measurements containing sub-wavelength information. The DL curves are reproduced from Fig.~2(g) of the main text.}
\label{figSZ}
\end{figure}

In Fig.~\ref{figSZ} we compare our results based on deep learning from Fig.~2(g) of the main text with corresponding results obtained using a linear multi-variable regression as digital decoder. The results suggest that in principle the latter is also capable of achieving deeply sub-wavelength resolution (which may (slightly) exceed the training data resolution). However, deeply sub-wavelength resolution is only observed at noise levels that are six or more orders of magnitude smaller. In other words, our ANN displays a remarkable robustness to noise. The linear multi-variable regression can only achieve good results at unrealistically low signal-to-noise ratios. Moreover, we note that for the ``no cavity'' case it does not manage to extract any information, and for the other cases it appears to saturate at resolutions about an order of magnitude worse than those achieved with deep learning at a noise magnitude that is six orders of magnitude higher. Finally, we note that a similar noise robustness of ANN decoders was also observed in Ref.~\cite{localiz_dynamic_environm_B}, albeit for a problem that did not consider any sub-wavelength information.

\subsection{Further Parameters that Impact the Localization Precision}

In Fig.~2(g) of the main text, we report the localization precision as a function of the measurement noise and the quality factor of the cavity. Here, we complement these results with three further parameters that were kept fixed in Fig.~2(g) but are now varied one at a time, while keeping the others fixed. Overall, these results further underline the absence of any fundamental wavelength-induced bound on the achievable localization precision. By choosing suitable parameters, the performance can be improved significantly and, as discussed in the introduction of the main text, even trivial setups (like unlabelled far-field plane wave approaches) can easily achieve resolutions well beyond the diffraction limit. An important insight is therefore that in reporting a physical mechanism to improve the achievable resolution, one should benchmark it against a reference case rather than merely reporting that one achieves sub-wavelength resolution.

\subsubsection{Training Data Resolution}

The finer the training data $\textbf{T}(X,f)$ resolves the allowed trajectory of the object, the more precisely the ANN should be able to approximate the mapping from measurement vector to object position. Beyond some point, however, the measurement noise exceeds the typical difference between two neighbouring sampling points such that they cannot be distinguished anymore. The parameter $T$ defines the training data resolution as illustrated in Fig.~\ref{figSY}(a). 

As expected, we observe in Fig.~\ref{figSY}(c) for the case of the cavity with $Q=263$ that lowering $T$ generally results in a higher localization error. We note that the achievable localization error can be considerably lower than the training data resolution for all considered values of $T$ except $T=9$. Moreover, we note that even with an extremely coarse training data resolution such as $T=17$ we achieve a clearly superior performance in the cavity with $Q=263$ than using an extremely fine training data resolution ($T=257$) in the case without cavity.

\subsubsection{Frequency Interval Sampling}

In principle, the finer a given frequency interval is sampled, the more information one can extract. This is generally true as long as the different samples are reasonably independent. The parameter $F$ defines the measurement's spectral sampling as illustrated in Fig.~\ref{figSY}(b). 
A rule of thumb is that frequency samples should be separated by at least $\Delta f_{\text{corr}} = f_0 / Q$ although this definition of spectral DoF is only approximative. The degree of independence can be evaluated more rigorously based on the singular value spectrum of the 2D training data matrix $\textbf{T}(X,f)$ (one dimension being different object positions, the other dimension being different frequency samples -- see Fig.~2(f) in the main text). Variables resulting from a random process are usually not perfectly independent but present a finite amount of correlations~\cite{del2019optimally_B,del2020optimal_B}. Therefore, although increasing the frequency interval sampling is initially expected to improve the localization accuracy, the marginal improvement from further increases of the frequency interval sampling eventually tends to zero.

As expected, we observe in Fig.~\ref{figSY}(d) for the case of the cavity with $Q=263$ that lowering $F$ generally results in a higher localization error. We note that the achievable localization error can be considerably below the training data resolution in all considered cases. Moreover, we note that even using an extremely low value of $F=9$ in the case of the cavity with $Q=263$ yields a localization error that can be orders of magnitude below that of the case without cavity using $F=257$.

\subsubsection{Object Size}

\begin{figure*}
\centering
\includegraphics [width =  \columnwidth]{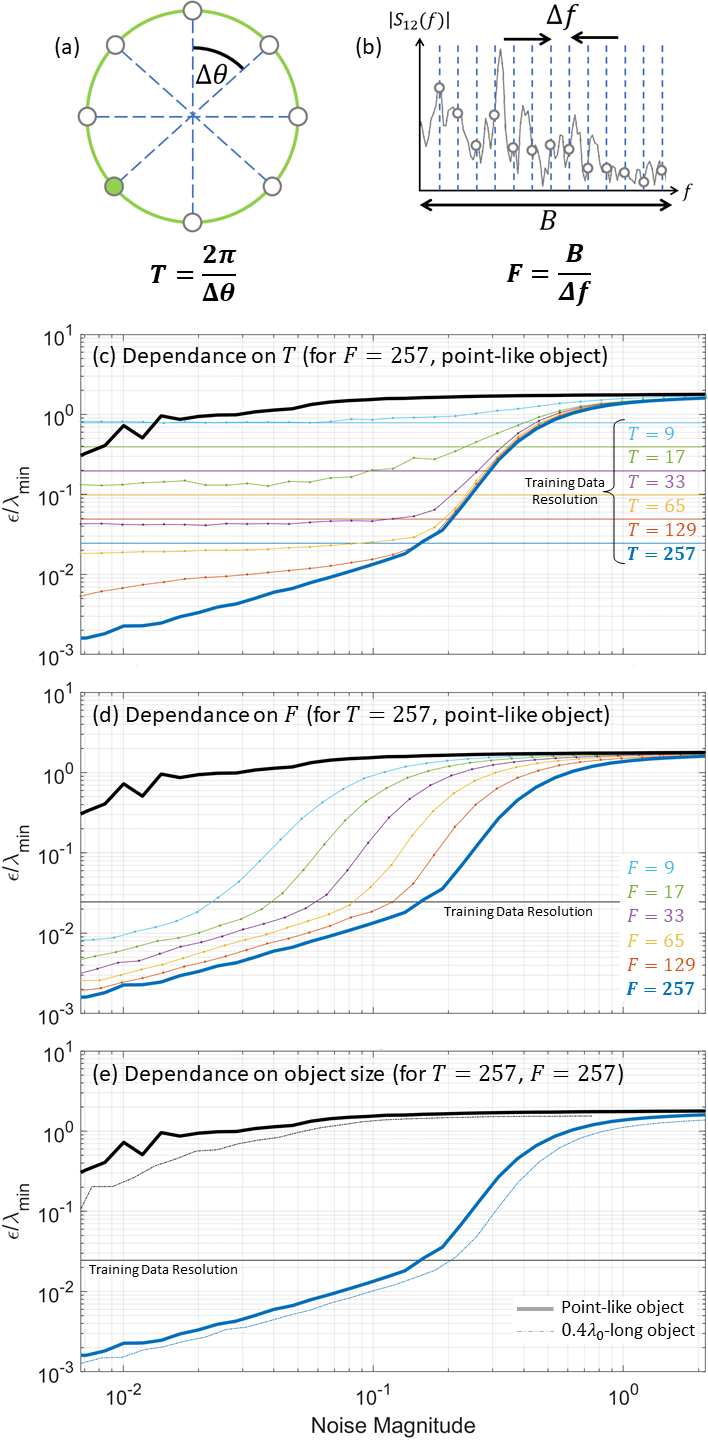}
\caption{Further parameters that impact the localization precision. 
(a) Definition of parameter $T$ to quantify the training data resolution. The allowed trajectory on which the object may move is divided into $T$ segments of equal length $R\Delta\theta$. 
(b) Definition of parameter $F$ to quantify the frequency interval sampling. A fixed interval $B$ is divided into $F$ segments of equal size $\Delta f$. 
(c) Dependence on $T$. The average localization error is plotted for the cavity with $Q = 263$ (Fig.~2(b) in the main text) for different training data resolutions $T$ (color-coded), keeping all other parameters fixed. Horizontal lines indicate the corresponding training data resolution in terms of the smallest utilized wavelength. For reference, the thick black curve representing the case without cavity (Fig.~2(a) in the main text) for $T=257$ is indicated. The black and blue thick curves are the same as the blue and red curves in Fig.~2(g) of the main text, respectively.
(d) Dependence~on~$F$. 
Same as (c) but with the roles of $T$ and $F$ inverted. All plotted curves correspond to the same training data accuracy (black horizontal line).
(e) Dependence on the object size. The average localization error is plotted for the point-like object considered thus far in the simulations (thick lines) as well as for a line object consisting of two scatterers $0.4\lambda_0$ apart (dash-dotted lines) is indicated, both for the case without cavity (black) and for the cavity with $Q=263$ (blue).}
\label{figSY}
\end{figure*}

The larger the object is, the more it scatters waves and the easier one should be able to localize it based on the limited amount of information contained in measurements with a given noise level, a given training data resolution and a given sampling of the considered frequency interval. In Fig.~2 of the main text, we considered the most difficult case: a non-resonant point-like object. In Fig.~\ref{figSY}(e), we compare these results with a larger line-like object composed of two point-like dipoles separated by $0.4 \lambda$. In principle, objects may also be effectively ``larger'' than others despite the same physical size due to being resonant or not, resulting in distinct scattering cross-sections.

As expected, we observe in Fig.~\ref{figSY}(e) that considering the larger object yields a small but notable improvement of the localization accuracy. The improvement is more pronounced for the case of the cavity with $Q=263$ than for the case without cavity.

\subsubsection{The Role of Signal Strength Enhancements}

\begin{figure*}[b]
\centering
\includegraphics [width =  15cm]{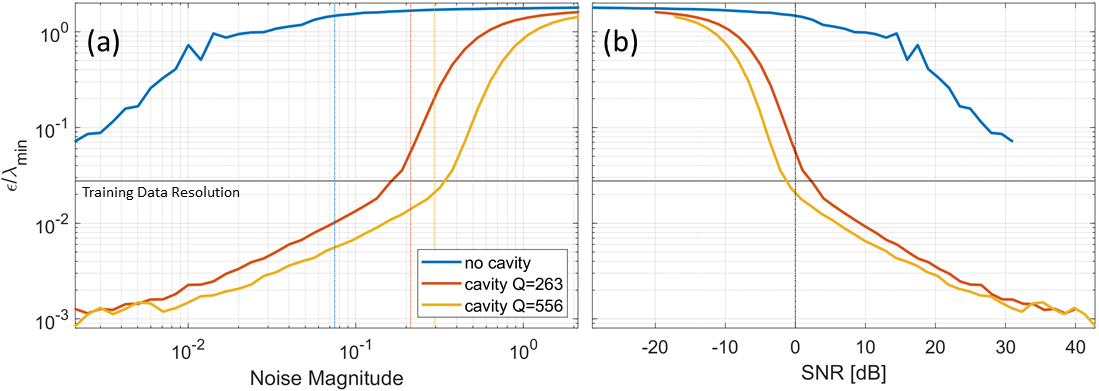}
\caption{Average localization error $\epsilon$ in terms of the smallest utilized wavelength $\lambda_{\text{min}}$ as a function of the \textit{absolute} (a) or \textit{relative} (b) magnitude of the measurement noise for the three cases considered in Fig.~2(a-c) of the main text. Vertical dashed lines indicate the corresponding signal magnitudes. The horizontal black line indicates the resolution of the training data. Note that (a) and Fig.~2(g) of the main text are identical.}
\label{figSX}
\end{figure*}

We clarified in the main text that the achievable resolution depends on various parameters such as the SNR, the dwell time in the interaction domain, and the number and independence of the utilized measurement modes. In order to clearly illustrate that the signal strength enhancement is not the only relevant factor in explaining the improved localization precision as the cavity's quality factor is increased, we contrast Fig.~2(g) from the main text, reproduced in Fig.~\ref{figSX}(a), which visualizes the dependence of $\epsilon$ on the \textit{absolute} magnitude of the measurement noise with the dependence of $\epsilon$ on the \textit{relative} magnitude of the measurement noise, that is, as function of the SNR, in Fig.~\ref{figSX}(b). In other words, the curves in Fig.~\ref{figSX}(b) are independent of any signal strength enhancement. The vertical dashed lines indicate zero SNR, hence they are distinct in (a) but coincide in (b). The differences between the three scenarios are somewhat reduced in (b) as compared to (a) since differences in signal strength are ignored, but the three curves continue to be significantly different, evidencing that the benefits of the cavity go beyond boosting the signal strength.

\clearpage
\providecommand{\noopsort}[1]{}\providecommand{\singleletter}[1]{#1}%

\end{document}